\documentclass[aps,preprint,tightenlines,showpacs,showkeys,nofootinbib]{revtex4}

\usepackage{graphicx}  
\usepackage{bm}  
\usepackage{amsmath}
\newcommand{\beq}{\begin{equation}}
\newcommand{\eeq}{\end{equation}}
\newcommand{\bea}{\begin{eqnarray}}
\newcommand{\eea}{\end{eqnarray}} 
\newcommand{\beqa}{\begin{eqnarray}}
\newcommand{\eeqa}{\end{eqnarray}} 
\newcommand{\nn}{\nonumber \\ }

\newcommand{\simlt}{\stackrel{<}{{}_\sim}}

\newcommand{\boma}[1]{\mbox{\boldmath{$#1$}}}
\def\vec#1{{\bf #1}}    
\newcommand{\fet}[1]{\mbox{\boldmath $#1$}}
\newcommand{\sing}{^1{\rm S}_0}
\newcommand{\trip}{^3{\rm S}_1}
\newcommand{\mpip}{M_\pi^{phys}}
\newcommand{\mpic}{M_\pi^{crit}}

\begin{document}
\title{More on the infrared renormalization group limit cycle in QCD}
\author{E. Epelbaum}\email{epelbaum@jlab.org}
\affiliation{Jefferson Laboratory, Theory Division, Newport News, 
VA 23606, USA}
\author{H.-W. Hammer}\email{hammer@itkp.uni-bonn.de}
\affiliation{Helmholtz-Institut f\"ur Strahlen- und Kernphysik (Theorie),
Universit\"at Bonn, Nu\ss allee 14-16, D-53115 Bonn, Germany}
\author{Ulf-G. Mei\ss ner}\email{meissner@itkp.uni-bonn.de}
\affiliation{Helmholtz-Institut f\"ur Strahlen- und Kernphysik (Theorie),
Universit\"at Bonn, Nu\ss allee 14-16, D-53115 Bonn, Germany
}\affiliation{
Institut f\"ur Kernphysik (Theorie), Forschungszentrum
J\"ulich, D-52425 J\"ulich, Germany}
\author{A. Nogga}\email{a.nogga@fz-juelich.de}
\affiliation{
Institut f\"ur Kernphysik (Theorie), Forschungszentrum
J\"ulich, D-52425 J\"ulich, Germany}

\date{\today}
\begin{abstract}
We present a detailed study of the recently conjectured 
infrared renormalization group limit cycle in QCD
using chiral effective field theory. It was conjectured that
small increases in the up and down quark masses can move
QCD to the critical trajectory for an infrared limit cycle in 
the three-nucleon system. At the critical quark masses,
the binding energies of the deuteron and its spin-singlet partner
are tuned to zero and the triton has infinitely many excited states
with an accumulation point at the three-nucleon threshold.
We exemplify three parameter sets where this effect occurs
at next-to-leading order in the chiral counting.
For one of them, we study the structure of the three-nucleon 
system in detail using both chiral and contact effective field 
theories. Furthermore, we investigate the matching of
the chiral and contact theories in the critical region and
calculate the influence of
the limit cycle on three-nucleon scattering observables.

\end{abstract}

\maketitle
\section{Introduction}

The renormalization group (RG) is an important tool
in many areas of physics.
Its applications range from critical phenomena in condensed matter physics
to the nonperturbative formulation of quantum field theories
for elementary particles \cite{Wilson:dy}. The requirement of the  
invariance of low-energy observables under changes of the 
ultraviolet cutoff $\Lambda$ generates a RG flow on the multidimensional 
space of coupling constants ${\bf g}$ for operators in the Lagrangian:
\begin{eqnarray}
\Lambda \frac{d}{d \Lambda} {\bf g} = \boma{\beta}({\bf g} ) \,.
\label{RGEq}
\end{eqnarray}
Depending on the physical application,
the RG flow can be considered in the infrared ($\Lambda\to 0$)
or in the ultraviolet ($\Lambda\to\infty$) limits.
The simplest solution to the RG equations is a fixed point ${\bf g}_*$ which
satisfies $\boma{\beta}({\bf g}_*) = 0$.
An important signature of an RG fixed point
is {\it scale invariance}: symmetry with respect to the
coordinate transformation ${\bf r} \to \lambda {\bf r}$
for any positive number $\lambda$.
This symmetry implies that dimensionless variables
scale as powers of the momentum scale.
Scale-invariant behavior at long distances, as in critical phenomena,
can be explained by RG flow to an infrared fixed point.
Scale-invariant behavior at short distances, as in
asymptotically-free field theories,
can be explained by RG flow to an ultraviolet fixed point.

However, RG equations can also exhibit more complicated solutions.
The possibility of RG flow to a limit cycle was proposed by
Wilson in 1971 \cite{Wilson:1970ag}.
A limit cycle is a 1-parameter family of coupling constants
${\bf g}_*(\theta)$ that is closed under the RG flow
and can be parametrized by an angle $0 < \theta < 2 \pi$.
The RG flow carries the system
around a complete orbit of the limit cycle every time
the ultraviolet cutoff $\Lambda$ increases by some factor $\lambda_0$.
One of the signatures of an RG limit cycle
is {\it discrete scale invariance}: symmetry with respect to the
coordinate transformation ${\bf r} \to \lambda_0^n {\bf r}$
only for integer values of $n$.

Before the fundamental theory of the strong interactions was known,
Wilson had suggested that limit cycles might be relevant to the 
high-energy behavior of the strong interactions of elementary particles 
\cite{Wilson:1970ag}.
However, Quantum Chromodynamics (QCD) and asymptotic freedom 
were discovered soon thereafter \cite{Gross-Politzer} 
and the high-energy behavior of the strong interactions
was explained by an ultraviolet fixed point in QCD.
The low-energy structure of the strong interactions, however, is 
much more complicated and not dominated by the fixed point.

The low-energy sector can be described by exploiting
the approximate chiral symmetry of QCD using
chiral effective field theory (EFT) methods 
\cite{Weinberg:1978kz,Gasser:1984gg,Bernard:1995dp}.
Chiral EFT is a powerful tool for analyzing the properties
of hadronic systems at low energies in a systematic and model-independent
way. It is formulated in an expansion around the chiral limit of QCD
which governs low-energy hadron structure and dynamics. In chiral
EFT's the quark mass dependence of the operators in the effective
Lagrangian is included explicitly. Over the past 15
years, considerable progress has been made in understanding the 
structure of the nuclear force in this framework 
\cite{Weinberg:1991um,Beane:2000fx,Bedaque:2002mn,Meissner:2004yy,
Epelbaum:2005pn}.

The quark mass dependence of the chiral nucleon-nucleon ($NN$)
interaction was 
studied with the primary aim to understand the chiral limit of nuclear 
physics \cite{Beane:2001bc,Beane:2002xf,Epelbaum:2002gb}.
While this question has not been fully resolved, these studies found that
the inverse scattering lengths in the $\trip$--$^3{\rm D}_1$ 
and $\sing$ channels 
both vanish if one extrapolates away from the physical limit
to slightly larger quark masses.\footnote{Due to the nuclear tensor
force, the $\trip$ and $^3{\rm D}_1$ channels are coupled. This mixing is
included in the chiral EFT calculations while it appears as a higher order 
effect in the contact EFT discussed below. For simplicity, we will only 
refer to the $\trip$ and $\sing$ partial waves in the following.}
Subsequently, it was pointed out that QCD is close to 
the critical trajectory for an infrared RG limit cycle in the 3-nucleon 
sector. This led to the conjecture, that QCD could 
be tuned to the critical trajectory by small changes in the up and down 
quark masses away from their physical values \cite{Braaten:2003eu}. 
The proximity of the physical quark masses to these critical values
explains the successful description of the low-energy three-nucleon 
problem in terms of zero-range forces between nucleons initiated
long ago \cite{STM57}.
The effective-field-theory formulation of this program
exhibits an ultraviolet RG limit cycle \cite{Bedaque:1999ve}.
The proximity of physical QCD to the critical trajectory
implies that the ultraviolet limit cycle of
Ref.~\cite{Bedaque:1999ve} is not just an artifact of the 
EFT but hints towards a limit cycle in QCD.

The connection between the limit cycle and three-nucleon
observables is established by the Efimov effect \cite{Efi71}
which occurs in the three-body sector of nonrelativistic particles
with a resonant short-range S-wave two-body interaction.
The strength of the interaction is governed by the S-wave scattering
length $a$. If $a$ is large and positive, there is a shallow
two-body bound state with binding momentum $\kappa=1/a$, if $a$ is large and 
negative there is a shallow virtual state characterized by the 
momentum scale $1/|a|$. Efimov showed that if $|a|$ is much larger than
the range $r_0$ of the interaction, there are shallow three-body bound states
whose number increases logarithmically with $|a|/r_0$.
In the  resonant limit $a \to \pm \infty$,
there are infinitely many shallow three-body bound states
with an accumulation point at the three-body scattering threshold.
If the particles are identical bosons,
the ratio of the binding energies of successive states
rapidly approaches the universal constant
$\lambda_0^2 \approx 515$.
Efimov also showed that low-energy three-body observables
for different values of $a$ are related by a
discrete scaling transformation in which $ a \to \lambda_0^n a$,
where $n$ is an integer, and lengths and energies are scaled by
the appropriate powers of $\lambda_0^n$ \cite{Efi71,Efi79}.
The mathematical connection between the Efimov effect and RG limit cycles
was first pointed out in Ref.~\cite{Albeverio:zi}.

The Efimov effect can also occur for fermions with
at least three distinct spin or isospin states and therefore
applies to nucleons as well.
The spin-singlet and  spin-triplet $np$ scattering lengths are
$a_{\sing} = -23.8$ fm and $a_{\trip} = 5.4$ fm.
They are both significantly larger than the effective range,
which is $r_0 =1.8$ fm in the spin-triplet channel. Efimov used this
observation  as the basis for a qualitative approach to the
three-nucleon problem \cite{Efi81}. A convenient implementation 
of this program is given by the so-called pionless or contact
EFT \cite{Bedaque:2002mn,Braaten:2004rn}. 
Nucleons are described as point particles with zero-range 
interactions whose strengths are adjusted
to reproduce the scattering lengths $a_{\trip}$ and $a_{\sing}$.
The effective range and higher order terms in the low-energy
expansions of the phase shifts are treated as perturbations.
This approach works well in nuclear few-body systems dominated
by momenta small compared to $M_\pi$. 
In the triton channel, the Efimov effect makes it necessary to 
include a three-body force at leading order in the power counting
\cite{Bedaque:1999ve}.
The three-body force can be fixed by using one three-body datum
as input. All other three-body observables can then be
predicted. The structure of this EFT is much simpler
than the chiral EFT and the computational effort
is considerably smaller. 
Since the contact EFT is based on an expansion around 
the limit of infinite scattering length, it is
particularly well suited to describe processes
governed by the large scattering length. As a consequence, the chiral
and contact EFT's can mutually complement each other. 
A first exploratory study of the infrared limit cycle in QCD 
in the contact EFT was carried out in Ref.~\cite{Braaten:2003eu}.
The quark mass dependence of the nucleon-nucleon scattering lengths from 
Ref.~\cite{Epelbaum:2002gb} was used as input in this calculation. 

In this paper, we study the possibility of an infrared limit cycle 
in QCD in more depth. We  use both chiral EFT and pionless EFT and
combine the strengths of both approaches. We use
the chiral EFT to calculate the bound state spectrum of the triton 
in the vicinity of the limit cycle and study how well it is approximated
by the contact EFT. Furthermore, we calculate various three-body scattering
observables in pionless EFT and illustrate how they are affected by the
limit cycle.

\section{Chiral Effective Field Theory}

The quark mass dependence of the chiral $NN$ interaction was calculated
to next-to-leading order (NLO) in the chiral counting 
in Refs.~\cite{Beane:2001bc,Beane:2002xf,Epelbaum:2002gb}.
At this order, the quark mass dependence is synonymous to
the pion mass dependence because of the Gell-Mann-Oakes-Renner relation:
\beq
M_\pi^2 = -(m_u + m_d) \langle 0 | \bar{u} u | 0 \rangle/F_\pi^2\,,
\label{eq:mq-mpi}
\eeq
where $\langle 0 | \bar{u} u | 0 \rangle \approx (-225 \mbox{ MeV})^3$
is the quark condensate. In the following, we will therefore refer
only to the pion mass dependence which is more convenient for 
nuclear applications and treat the pion mass as a parameter that can be
varied by adjusting the values of the quark masses.
In the work of Refs.~\cite{Beane:2001bc,Beane:2002xf,Epelbaum:2002gb},
it was found that the scattering lengths in the $\trip$--$^3{\rm D}_1$ 
and $\sing$ channels both vanish in the pion mass region around 200 MeV.

In this work, we study the structure of the nuclear three-body
system in this region of pion masses around 200 MeV.
We will use the chiral $NN$ potential constructed from EFT 
using the method of unitary transformations \cite{Epelbaum:2002gb}.
To next-to-leading order (NLO) in the chiral power counting,
this potential can be written as:
\beq
V_{\rm NLO} = V^{\rm OPE} + V^{\rm TPE} + V^{\rm cont}\,,
\eeq
where $V^{\rm OPE}$, $ V^{\rm TPE}$, and  $V^{\rm cont}$
refer to the one-pion exchange, two-pion exchange, and contact potentials,
respectively. 
They are given by the expressions:
\beqa
\label{potfin}
V^{\rm OPE} &=& - \frac{1}{4} \frac{g_A^2}{F_\pi^2} \left( 1 + 2 \Delta
- \frac{4 M_\pi^2}{g_A} \bar d_{18} \right) 
\, \fet \tau_1 \cdot \fet \tau_2 \, \frac{(\vec \sigma_1 \cdot \vec q \,) 
( \vec \sigma_2 \cdot \vec q\,)}
{\vec q\, ^2 + M_\pi^2} ~, 
\eeqa
\beqa
V^{\rm TPE} &=& - \frac{ \fet{\tau}_1 \cdot \fet{\tau}_2 }{384 \pi^2 
F_\pi^4}\,\biggl\{
L(q) \, \biggl[4 M_\pi^2 (5g_A^4 - 4g_A^2 -1) + \vec q\, ^2 (23g_A^4 - 
10g_A^2 -1) 
+ \frac{48 g_A^4 M_\pi^4}{4 M_\pi^2 + \vec q\, ^2} \biggr] \nn
&& {} \mbox{\hskip 2 true cm} + \vec q\, ^2 \, \ln \frac{M_\pi}{\mpip} 
\, (23g_A^4 - 10g_A^2 -1) \biggr\} \nn
&&{} - \frac{3 g_A^4}{64 \pi^2 F_\pi^4} \,\left( L(q) + 
\ln \frac{M_\pi}{\mpip} \right)\, \biggl\{
\vec{\sigma}_1 \cdot\vec{q}\,\vec{\sigma}_2\cdot\vec{q} - \vec q\,^2 \, 
\vec{\sigma}_1 \cdot\vec{\sigma}_2 \biggr\}~,
\label{potfinTPE} 
\eeqa
\beqa
 V^{\rm cont} &=& \bar C_S + \bar C_T (\vec \sigma_1 \cdot \vec \sigma_2 ) 
+ M_\pi^2 \, \left( \bar D_S - \frac{3 g_A^2}{32 \pi^2 F_\pi^4} 
( 8 F_\pi^2  C_T- 5 g_A^2 + 2) 
\ln \frac{M_\pi}{\mpip} \right) \nn
&& {}+ M_\pi^2 \left( \bar D_T - \frac{3 g_A^2}{64 \pi^2 F_\pi^4} 
( 16 F_\pi^2  C_T- 5 g_A^2 + 2) 
\ln \frac{M_\pi}{\mpip} \right)\,
(\vec \sigma_1 \cdot \vec \sigma_2 ) \nn
&& {} + C_1 {\vec q \,}^2 
+ C_2 {\vec k \,}^2 + ( C_3 {\vec q \, }^2 + C_4 {\vec k \,}^2 ) 
( \vec \sigma_1 \cdot \vec \sigma_2 ) \nn
&& {} + i C_5 \frac{ \vec \sigma_1 + \vec \sigma_2}{2} 
\cdot ( \vec k \times \vec q \, ) 
+ C_6 ( \vec q \cdot \vec \sigma_1 ) 
( \vec q \cdot \vec \sigma_2 ) + C_7 ( \vec k \cdot \vec \sigma_1 )
( \vec k \cdot \vec \sigma_2 ) \,,
\label{potfincont}
\eeqa
with $g_A$ and $F_\pi$ the physical values of the nucleon axial coupling and 
pion decay constant, respectively and $\bar d_{18}$ a low-energy constant 
related to the Goldberger-Treiman discrepancy \cite{Epelbaum:2002gb}.
The symbols $\sigma_i$ ($\fet \tau_i$), $i=1,2$ indicate the spin (isospin) 
operators for particle $i$ and the $C_{1,...,7}$, $\bar C_{S,T}$, 
and $\bar D_{S,T}$ are low-energy constants (LEC's)
to be determined from fits to nucleon-nucleon data.
Further, $\vec q$ denotes the momentum transfer of the nucleon, 
i.e.~$\vec q = \vec p ' - \vec p$, where $\vec p '$ and $\vec p$ 
are final and initial nucleon momenta, while $\vec k = (\vec p ' + \vec p)/2$.
Here and in what follows  we denote the value of the variable pion mass by 
$M_\pi$ in order to distinguish it from the physical value denoted 
by $\mpip=139.6$ MeV. Furthermore, 
\beq
L(q) \equiv L(| \vec q \,|) 
= \frac{\sqrt{4 M_\pi^2 + \vec q\, ^2}}{|\vec q \,|} \, 
\ln \frac{ \sqrt{4 M_\pi^2 + \vec q\, ^2}
+ | \vec q \, |}{2 M_\pi}\, ,
\eeq
and $\Delta$ represents the relative shift in the ratio $g_A/F_\pi$ compared 
to its physical value:
\beqa
\Delta &\equiv& \frac{\left(g_A/F_\pi\right)_{M_\pi}-\left(
g_A/F_\pi\right)_{\mpip}}{
\left(g_A/F_\pi\right)_{\mpip}} \nn
&=& \left( \frac{g_A^2 }{16 \pi^2 F_\pi^2} - \frac{4 }{g_A}
\bar{d}_{16} + \frac{1}{16 \pi^2 F_\pi^2} \bar{l}_4 \right) 
\left((\mpip)^2 - M_\pi^2\right) 
- \frac{g_A^2 M_\pi^2}{4 \pi^2 F_\pi^2} \ln \frac{M_\pi}{\mpip} 
\, ,
\label{deltaCL}
\eeqa
where the low-energy constants $\bar{d}_{16}$ and $\bar{l}_4$
are defined as in Ref.~\cite{Epelbaum:2002gb}.
Note that in the TPEP,
we only take into account the explicit $M_\pi$--dependence 
and use the physical values for $g_A$ and $F_\pi$. This is sufficient at NLO 
since any  shift in $g_A$ and $F_\pi$ for a different value of $M_\pi$ in 
the TPE  is a N$^4$LO effect. 
We also incorporate the leading isospin--breaking corrections 
due to the pion mass difference in the one--pion exchange potential
\cite{vanKolck:1996rm}
but do not consider an independent variation of $m_u - m_d$.

The constants 
$\bar C_{S,T}$ and $\bar D_{S,T}$ are related to the $C_{S,T}$
used in \cite{EGM2} via
\beq
C_{S,T} = \bar C_{S,T} + (\mpip)^2 \bar D_{S,T}\,.
\label{eq:CDmpi}
\eeq
Note further that the short--range terms of the type $M_\pi^2 
\ln M_\pi$
in Eq.~(\ref{potfincont}) result from the two--pion exchange as well as from 
the renormalization 
of the leading--order contact forces by pion loops. 
It is important to stress that renormalization of the LECs  
$C_S$, $C_T$, $C_{1, \ldots 7}$ due to pion loops does not depend on the pion 
mass and  thus is of no relevance for this work. The  
potential $V (\vec p \, ', \, \vec p \, )$ is multiplied by the regulating 
functions $f_{\rm R} (| \vec p \, |)$, $f_{\rm R} (| \vec p \, ' |)$ 
in order to cut off the large momentum components in the 
Lippmann--Schwinger equation.
In this study, we use the same exponential function $f_R (| \vec p \, |) = 
\exp [ -\vec p \, ^4/\Lambda^4]$
as in Ref.~\cite{3Nno3NF} and
restrict ourselves to the cut--off $\Lambda = 540$ MeV\footnote{The cutoff 
dependence is mild in the very low energy regime, in which we 
are interested here. For a discussion of larger cutoffs, see 
Ref.~\cite{Nogga:2005hy}.}. 

In principle, these equations determine the pion mass dependence of the
chiral $NN$ potential uniquely. However, the extrapolation away from the 
physical pion mass generates errors. The dominating source are the
constants  $\bar C_{S,T}$ and $\bar D_{S,T}$ which cannot be determined
independently from fits to data at the physical pion mass.
A smaller effect is due to the error in the LEC $\bar d_{16}$ 
which is enhanced in 
Eq.~(\ref{deltaCL})
as one moves away from the physical pion mass. Both effects generate
increasing uncertainties as one extrapolates away from the physical point.
Note also that we do not include explicit $\Delta(1232)$ degrees of freedom.
It would be interesting to see if and how our results would be modified in
a theory with explicit  $\Delta$'s \cite{Bedaque:2002mn}.

Following Ref.~\cite{Epelbaum:2002gb}, we use $\bar d_{16}=-1.23$ GeV$^{-2}$ 
which is the average of three values given in Ref.~\cite{Fettes:2000fd} 
(See also Ref.~\cite{Fettes:1999wp}).
The LEC $\bar d_{18} = -0.97$ GeV$^{-2}$ is determined from the observed 
value of the Goldberger--Treiman discrepancy while 
the two remaining constants $\bar D_{S}$ and $\bar D_{T}$
are unknown. The size of these two constants can be
constrained from naturalness arguments. In Ref.~\cite{Epelbaum:2002gb},
it was argued that the 
corresponding dimensionless constants $F_\pi^2 \Lambda_\chi^2 \bar D_{S,T}$ 
can be expected to satisfy the bounds:
\beq
-3 \leq F_\pi^2 \Lambda_\chi^2 \bar D_{S,T} \leq 3\,,
\label{eq:STbound}
\eeq
where $\Lambda_\chi \simeq$ 1 GeV is the chiral symmetry breaking scale.
We note that Refs.~\cite{Beane:2001bc,Beane:2002xf} do allow for a larger 
variation of these LEC's. The LEC's that we are going to choose below 
are within our more restrictive range. Therefore, the discussion 
on the appropriate scaling of the $D_{S,T}$ is not relevant for this study.  
Furthermore,
these bounds are in agreement with resonance saturation estimates and
similar conditions are obeyed by the known constants \cite{Epelbaum:2001fm}.
For the constants $C_{S,T}$, e.g., we find 
$C_{S}=-120.8$ GeV$^{-2}$ and $C_{T}=1.8$ GeV$^{-2}$ corresponding 
to the dimensionless coefficients $F_\pi^2 C_{S}$ = $-$1.03 and 
$F_\pi^2 C_{T}$ = 0.02, respectively. The unnaturally small value
of $F_\pi^2 C_{T}$ is a consequence of the approximate Wigner
SU(4) symmetry. 
(For a discussion of this issue in the pionless
EFT, see Ref.~\cite{Mehen:1999qs}).

The ranges from Eq.~(\ref{eq:STbound}) were
used to estimate the extrapolation errors of two-nucleon observables
like the deuteron binding energy and the spin-singlet and 
spin-triplet scattering
lengths in Ref.~\cite{Epelbaum:2002gb}. In the exploratory study of
the three-nucleon system \cite{Braaten:2003eu}, the mean values of these 
error bands were used as input for the three-body calculations in the 
contact EFT. 
Even though both scattering lengths were large for the mean values,
they did not become infinite at the same value of the pion mass and 
there was no exact limit cycle for this choice of parameters.

Here we take a different approach and search for sets of values
for $\bar D_{S}$ and $\bar D_{T}$ that lie within the bound 
given by Eq.~(\ref{eq:STbound}) and cause the spin-singlet and 
spin-triplet scattering lengths to become infinite at the same value of 
the pion mass. For this purpose, it is more 
convenient to use the partial wave projected constants
\beqa
\bar D_{\trip} &=& 4\pi \left(\bar D_{S} + \bar D_{T}\right)\,,
\nonumber \\
\bar D_{\sing} &=& 4\pi \left(\bar D_{S} -3 \bar D_{T}\right)\,.
\eeqa
The dimensionless constants $\alpha_{\sing}$ and $\alpha_{\trip}$
defined as 
\beq
\alpha_{\sing}=F_\pi^2 \Lambda_\chi^2 \bar D_{\sing}/(16 \pi) \qquad
\mbox{and} \qquad
\alpha_{\trip}=F_\pi^2 \Lambda_\chi^2 \bar D_{\trip}/(8 \pi )\,,
\eeq
then satisfy the same bound as in Eq.~(\ref{eq:STbound}), i.e.:
\beqa
-3 &\leq& \alpha_{\trip} \leq 3\nonumber\,, \\
-3 &\leq& \alpha_{\sing} \leq 3\,.
\label{eq:31bound}
\eeqa
The values of the parameters $\alpha_{\trip}$ and $\alpha_{\sing}$
can be chosen independently anywhere in the above intervals without
violating naturalness and without affecting physics at the physical 
value of the pion mass.

The values of the pion mass where the inverse scattering 
lengths in the spin-triplet and spin-singlet channels
vanish simultaneously are called critical pion masses $\mpic$.
At NLO in the chiral counting,
it is possible to find parameter sets with critical
pion masses in the range 175 MeV $\simlt \mpic \simlt$ 205 MeV.
For example, for the three exemplifying values $\alpha_{\trip}
=\pm 2.5$ and 0.0, we obtain the following critical parameter sets:
\begin{itemize}
\item[(a)] $\alpha_{\trip}=-2.5$ and $\alpha_{\sing}=2.138598$ 
$\qquad\Longrightarrow\qquad$ $\mpic = 197.8577$ MeV,
\item[(b)] $\alpha_{\trip}=0.0$ and $\alpha_{\sing}=1.955709$
$\qquad\Longrightarrow\qquad$ $\mpic = 186.3276$ MeV,
\item[(c)] $\alpha_{\trip}=2.5$ and $\alpha_{\sing}=1.776665$ 
$\qquad\Longrightarrow\qquad$ $\mpic = 179.0417$ MeV.
\end{itemize}
We note that it is unlikely that physical QCD will correspond 
to any of the solutions (a)--(c). 
However, in Ref.~\cite{Braaten:2003eu} it was conjectured
that one should be able to reach the critical point by varying  
the up- and down-quark masses $m_u$ and $m_d$ independently
because the spin-triplet and spin-singlet channels have different isospin.
A more detailed investigation is needed in order to test this conjecture. 
However, many aspects of the limit cycle are universal and do not depend
on the exact parameter values \cite{Braaten:2004rn}. 
Therefore, we study the structure of the 
three-nucleon system near the critical pion mass for
solution (a) in the remainder of this paper in 
more detail. The universal aspects of the three-body
observables do not depend on the details of the solution we choose.

\begin{figure}[tb]
\centerline{\includegraphics*[width=9cm,angle=0,clip=true]{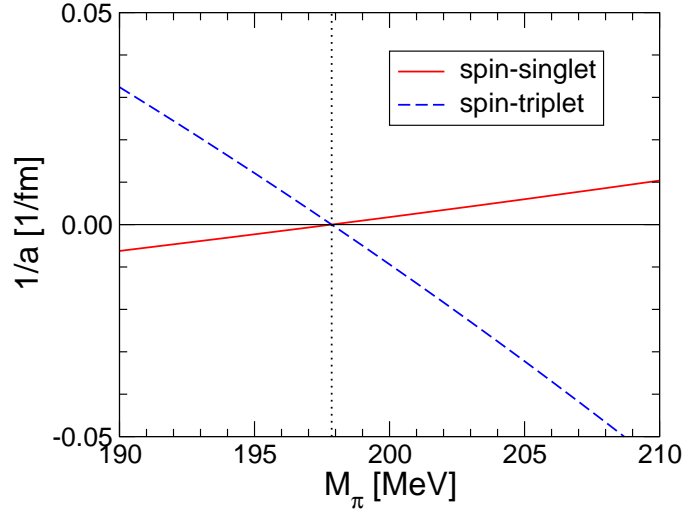}}
\caption{Inverse of the S-wave scattering lengths in the 
spin-triplet and spin-singlet nucleon-nucleon channels 
for solution (a) as a function of the pion mass $M_\pi$.
The vertical dotted line indicates the critical pion mass $\mpic$.}
\label{fig:asat}
\end{figure}

The inverse scattering lengths in the spin-triplet and  spin-singlet 
channels  in the vicinity of the limit cycle
for solution (a) are shown in Fig.~\ref{fig:asat}.
As promised,  the inverse scattering lengths vanish at the critical
value of the pion mass $\mpic = 197.8577$ MeV.
For pion masses below the critical value, the spin-triplet
scattering length is positive and the deuteron is bound. 
As the inverse spin-triplet scattering length decreases, 
the deuteron becomes more and more
shallow and finally becomes unbound at the critical mass. Above the critical 
pion mass the deuteron exists as a shallow virtual state. 
In the spin-singlet channel, the situation is reversed: the 
\lq\lq spin-singlet deuteron'' is a virtual state below the critical pion 
mass and becomes bound above. 
The pion mass dependence of the two scattering lengths shown
in Fig.~\ref{fig:asat} will be used as
input for the calculations in the contact EFT in the next section.

\begin{figure}[tb]
\centerline{\includegraphics*[width=9cm,angle=0,clip=true]{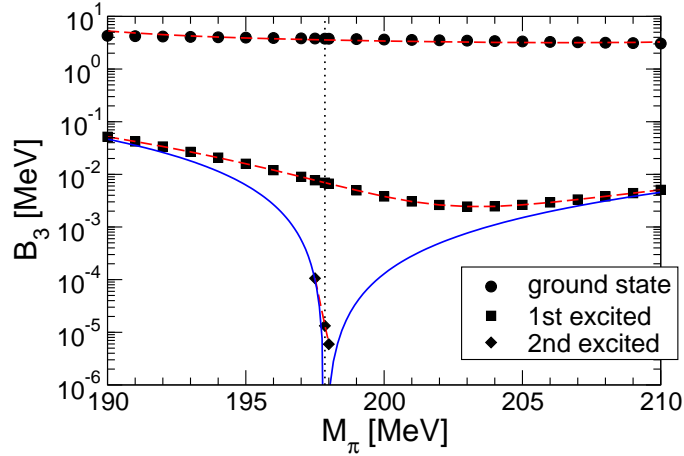}}
\caption{Binding energies $B_3$ of the triton ground state and the first two
excited states as function of the pion mass $M_\pi$.
The circles (ground state), squares (first excited state), and
diamonds (second excited state) give the chiral EFT result, while
the dashed lines are calculations in the pionless theory.
The vertical dotted line indicates the critical pion mass $\mpic$.
The thresholds for the three-body states are
given by the solid lines.
}
\label{fig:bind3}
\end{figure}

{}From the solution of the Faddeev equations with 
solution (a) for the $NN$ potential, 
we obtain the binding energies of the triton and the first
two excited states in the vicinity of the limit cycle
(See Ref.~\cite{Epelbaum:2002vt} for details).
The binding energies are given in Fig.~\ref{fig:bind3}
by the circles (ground state), squares (first excited state),
and diamonds (second excited state). The solid lines indicate the
neutron-deuteron ($M_\pi \leq \mpic$) and 
neutron-spin-singlet-deuteron ($M_\pi \geq \mpic$) thresholds
where the three-body states become unstable. Directly at 
the critical mass, these thresholds coincide with the three-body
threshold and the triton has infinitely many excited states.
The dashed lines are calculations in the pionless theory and
will be discussed in detail below.
The binding energy of the triton ground state
varies only weakly over the whole range of pion masses and is about 
one half of the physical value at the critical point. The excited states are
influenced by the thresholds and vary much more strongly.

In the remainder of this subsection,
we calculate the expectation values of the $2N$ and $3N$ kinetic energies
and some properties of the $2N$ and $3N$ wave functions. 
While these quantities are technically not observables, they shed some 
light on the structure of the three-body states.

\begin{figure}[tb]
\centerline{\includegraphics*[width=9cm,angle=0,clip=true]{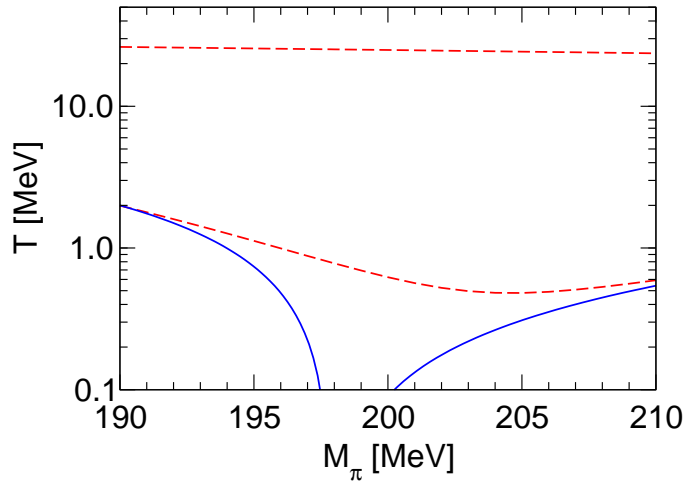}}
\caption{Kinetic energies of the triton ground state and the first
excited state indicated by the dashed lines compared to the kinetic energies 
of the deuteron ($M_{\pi} < M_{\pi}^{crit}$) 
and the spin-singlet deuteron ($M_{\pi} > M_{\pi}^{crit}$)
indicated by the solid lines as function of pion mass $M_\pi$.
}
\label{fig:ekin}
\end{figure}

In Fig.~\ref{fig:ekin}, we show the expectation values of the 
kinetic energy for the triton ground and first excited states
and for the two-nucleon states as a function of the pion mass $M_\pi$.
All expectation values are evaluated in the rest frame of the corresponding
states. The triton ground state kinetic energy stays fairly constant
as $M_\pi$ is varied. The kinetic
energy of the first excited state, however, approaches the kinetic
energy of the two-nucleon bound state (the deuteron for $M_{\pi} < 
M_{\pi}^{crit}$ and the spin-singlet deuteron for $M_{\pi} > M_{\pi}^{crit}$)
near the value of the pion mass where the triton excited state enters 
from the $2N$-$N$ continuum. As a consequence, the third particle 
is essentially at rest in this region. 
This behavior indicates that the first excited state has a
$2N$-$N$ cluster structure close to the $2N$-$N$ threshold. 
A similar observation 
applies to the second excited state which is not shown in the figure.

\begin{figure}[tb]
\centerline{\includegraphics*[width=9cm,angle=0,clip=true]{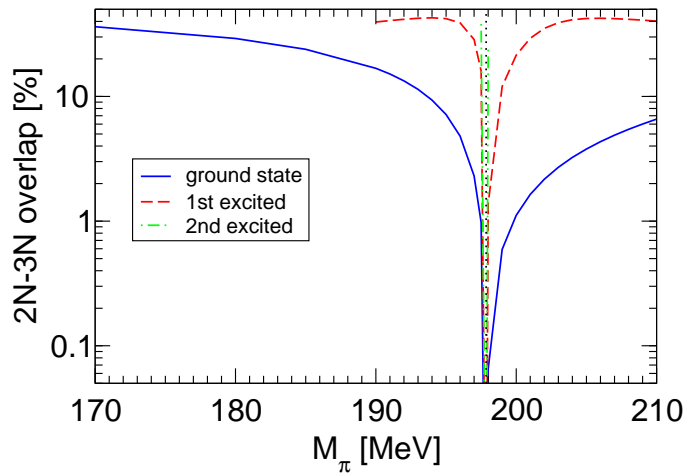}}
\caption{$2N$-$3N$ wave function overlap for the triton ground state 
and the first two excited states as function of pion mass $M_\pi$.
The vertical dotted line indicates the critical pion mass $\mpic$.}
\label{fig:overlap}
\end{figure}

The $2N$-$3N$ wave function overlap for 
the triton ground state and the first two 
excited states is shown in  Fig.~\ref{fig:overlap}
as function of pion mass $M_\pi$.
The overlap rapidly approaches zero as the pion mass reaches the
critical value. This is a consequence of 
increasing the size of the deuteron ($M_\pi \leq \mpic$) and 
spin-singlet-deuteron ($M_\pi \geq \mpic$) as the critical point is 
approached from below and above, respectively.
Directly at the critical point, they are both infinitely large.
While new shallow  three-body states appear at values of $B_3$ that form a 
geometric series and differ by factors of $\lambda_0^2=515$ as $\mpic$ is 
approached, the size of a given three-body state remains finite
(cf. Fig.~\ref{fig:bind3}). Consequently, the overlap
between $2N$ and $3N$ wave functions vanishes at the critical point.

We have also calculated the probability for the D-wave and mixed symmetry 
states in the triton ground state and the first two excited states
as function of pion mass $M_\pi$. These results will be discussed below 
in the context of the contact EFT.

\section{Contact Effective Field Theory}

In principle, one could calculate all three-nucleon observables 
in the vicinity of the critical pion mass in the chiral EFT. However,
the computational effort is significant, even in the 3N system. 
Moreover, the calculations become increasingly difficult 
for the shallow excited states near the limit cycle. 
The physics near the limit cycle, however, can also be described
in the contact EFT for large scattering length. This theory is formulated 
in an expansion around the limit of two-body bound states at threshold
corresponding to vanishing inverse scattering lengths. It is much
simpler than the chiral EFT and does not contain pion degrees of
freedom. Three-body calculations can typically be 
carried out on a personal computer. Therefore, it is an ideal tool to 
calculate physical observables in
the critical region where the scattering lengths are large.
We note that the contact
EFT is universal and does not have the explicit pion mass dependence. 
When the pion mass dependence of the spin-triplet and spin-singlet scattering 
lengths (cf.~Fig.~\ref{fig:asat})
as well as one three-body observable (cf.~Fig.~\ref{fig:bind3})
are taken from the chiral EFT calculation, the pion mass dependence of
other three-body observables can be calculated. Therefore,
it complements the chiral EFT study from the previous section.

For practical purposes, it is convenient to write down this theory
in the Lagrangian formalism using so-called \lq\lq dibaryon'' fields.
In our case, we need two dibaryon fields:
(i) a field $t_i$ with
spin (isospin) 1 (0) representing two nucleons interacting in the $^3 S_1$
channel (the deuteron) and
(ii) a field $s_a$ with
spin (isospin) 0 (1) representing two nucleons interacting in the $^1 S_0$
channel \cite{Bedaque:1999ve}:
\beqa
{\cal L}_{contact}&=&N^\dagger \Big(i\partial_t +\frac{\vec{\nabla}^2}
{2M}\Big)N
+ \frac{g_t}{2} t^\dagger_i t_i +\frac{g_s}{2} s^\dagger_a s_a
- \frac{g_t}{2}\Big( t^\dagger_i N^T \tau_2 \sigma_i
\sigma_2 N +h.c.\Big) \nonumber \\
& &-\frac{g_s}{2}\Big(s^\dagger_a N^T \sigma_2 \tau_a \tau_2 N +h.c.\Big) 
- \frac{2MH}{\Lambda^2}  N^\dagger \Big[ g_t^2
(t_i \sigma_i)^\dagger (t_j\sigma_j) \nonumber \\
 && \qquad \qquad\qquad +\frac{g_t g_s}{3} \left( (t_i\sigma_i)^\dagger
(s_a \tau_a) + h.c. \right) 
+ g_s^2 (s_a \tau_a)^\dagger (s_b \tau_b) \Big] N\,,
\label{lagd}
\eeqa
where $i,j$ are spin and $a,b$ are isospin indices,
$M$ is the nucleon mass, and $g_t$, $g_s$, and $H$ are the bare 
coupling constants. The $\sigma_i\, (\tau_a)$ are Pauli matrices
acting in spin (isospin) space.
In the two-body sector, the exact solution of the field theory
can be obtained analytically \cite{Kaplan:1996nv}.
Renormalization can be implemented by adjusting the two-body
coupling constants $g_t(\Lambda)$ and $g_s(\Lambda)$ as a function of
the ultraviolet momentum cutoff $\Lambda$ such that the 
spin-triplet and spin-singlet scattering lengths have the desired values.
Other two-body observables are then independent of $\Lambda$
and have the appropriate values up to corrections of order $r_0/|a|$
and $r_0 \sqrt{M|E|}$ where $r_0$ is the range of the interaction and 
$E$ the typical energy.

In the three-body sector, the nonperturbative solution of the field theory
can be obtained by solving generalized Skorniakov-Ter-Martirosian
integral equations including a three-body force
numerically. These integral equations have unique 
solutions only in the presence of an ultraviolet cutoff $\Lambda$.
The resulting predictions for three-body observables,
although finite, depend on the cutoff and are
periodic functions of $\ln(\Lambda)$ with period $\pi/s_0$ where
$s_0 = 1.0062378...$ is a transcendental number.
In Ref.~\cite{Bedaque:1999ve} it was shown that the quantum field theory
could be fully renormalized to remove the residual dependence
on $\Lambda$ in the three-body sector by adding a three-body interaction term
to the Lagrangian density in (\ref{lagd}).
The dependence of three-body observables on the cutoff
decreases like $1/\Lambda^2$ if the three-body coupling constant
has the form \cite{Bedaque:1998kgkm}
\begin{eqnarray}
H(\Lambda) = \frac{\cos[s_0 \ln(\Lambda/\Lambda_*) + \arctan(s_0)]}
{\cos[s_0 \ln(\Lambda/\Lambda_*) - \arctan(s_0)]}\,,
\label{Hlc}
\end{eqnarray}
for some value of $\Lambda_*$.
With this renormalization, three-body observables have well-defined
limits as $\Lambda \to \infty$, but they depend on
the parameter $\Lambda_*$ introduced by dimensional transmutation.
Since $H(\Lambda)$ is a periodic function of $\ln(\Lambda)$,
the renormalization of the field theory involves
an ultraviolet limit cycle. This EFT has succesfully been applied
to various nuclear three-body observables \cite{Bedaque:2002mn}.
Higher order corrections can be calculated as well 
\cite{Hammer:2001gh,Bedaque:tritn2lo,Afnan:2003bs,Griesshammer:2004pe} 
but are suppressed
by $r_0/|a|$ and $r_0 \sqrt{M|E|}$ near the exact limit cycle.
(For a recent formal study of the corrections in repulsive
partial waves, see Ref.~\cite{Birse:2005pm}.)

\begin{figure}[tb]
\centerline{\includegraphics*[width=9cm,angle=0,clip=true]{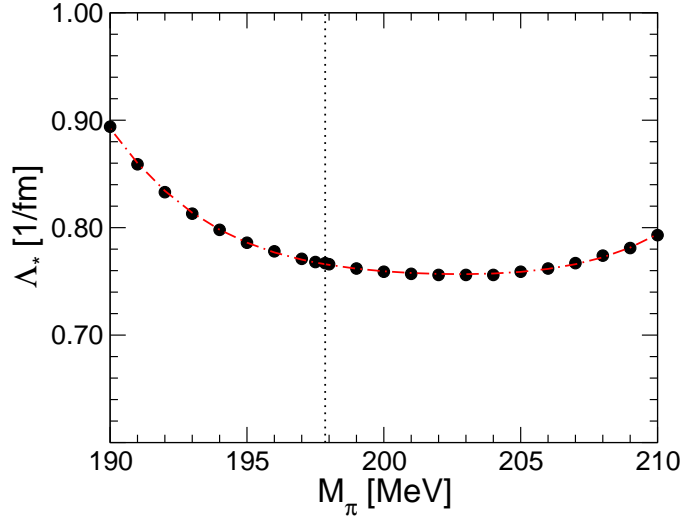}}
\caption{The three-body parameter $\Lambda_*$ determined from matching the 
first excited state of the triton as a function of the 
pion mass $M_\pi$ (circles). The dash-dotted line is a fourth order
polynomial fit while the vertical dotted line indicates the critical pion 
mass $\mpic$.}
\label{fig:Ls}
\end{figure}

As discussed above, we take the
pion mass dependence of the spin-singlet and spin-triplet
scattering lengths from the chiral EFT calculation (cf.~Fig.~\ref{fig:asat}). 
The pion mass dependence of the three-body
parameter $\Lambda_*$ can be determined from matching the energy of the
triton ground state or one of its excited states (cf.~Fig.~\ref{fig:bind3}).
Since the higher-order effects are generally smaller for the shallower states,
we match to the first excited state.
The pion mass dependence of all other three-body observables can then 
be predicted to leading order in the power counting.
In Fig.~\ref{fig:Ls}, the circles give the value of  $\Lambda_*$
obtained from the matching procedure as a function of the pion mass.
The pion mass dependence of $\Lambda_*$ is well described by a
fourth order polynomial fit indicated by the dash-dotted line.
For pion masses between 190 and 210 MeV, the parameter  $\Lambda_*$
varies smoothly by about 15\%. 
The 15\%  variation in $\Lambda_*$ corresponds to a variation of the triton 
ground state energy by about 25\% over the same range of pion masses.
In the previous work of Ref.~\cite{Braaten:2003eu}, the value of 
$\Lambda_*$ was approximated by a constant since no three-body
calculations with chiral potentials were available.

We are now in the position to calculate the pion mass dependence of the 
other three-nucleon observables.
The results for triton ground state and second excited state 
are compared to the chiral EFT results in Fig~\ref{fig:bind3}.
The circles (ground state), squares (first excited state), and
diamonds (second excited state) give the chiral EFT result, while
the dashed lines are calculations in the contact EFT.
We find good agreement between the chiral and contact EFT calculations
which is generally better for the shallower states.
The typical error is about 10\% for the ground state and below 1\% for 
the second excited state. The first excited state is reproduced 
exactly because of the matching procedure. 

The pionless theory can also be used to calculate the binding
energies of the next few excited states in the critical region
where it is very difficult to perform calculations in the chiral
EFT. Our results for the binding energies of the first 6 excited states
at the critical pion mass are compared to the results from chiral EFT
in Table \ref{tab1}. The second and 4th columns show the binding energies 
and ratios
\renewcommand{\arraystretch}{1.2}
\begin{table}[t]
\begin{center}
\begin{tabular}{|c|c|c|c|c|}
\hline\hline
$n$ & $B_3^{(n)}$ [MeV] (chir.) & $B_3^{(n)}$ [MeV] (cont.) 
& $B_3^{(n-1)}/B_3^{(n)}$ (chir.)& $B_3^{(n-1)}/B_3^{(n)}$ (cont.)
\\ \hline
$\quad -1 \quad$ & & $1.8437\times 10^{3}$  &  & 515.0 \\
$\quad 0 \quad$ & 3.7736 & 3.5798 &  & 515.0 \\
$\quad 1 \quad$ & $6.9504\times 10^{-3}$ & $6.9504\times 10^{-3}$ &  542.9
& 515.0 \\  
$\quad 2 \quad$ &  $1.3287\times 10^{-5}$&  $1.3495\times 10^{-5}$ & 523.1
& 515.0\\
$\quad 3 \quad$ & -- & $2.6202\times 10^{-8}$ &  -- & 515.0\\
$\quad 4 \quad$ &  -- & $5.0874\times 10^{-11}$ & -- &  515.0\\
$\quad 5 \quad$ &  -- & $9.8779\times 10^{-14}$ &  -- & 515.0\\
$\quad 6 \quad$ &  -- & $1.9179\times 10^{-16}$ &  -- & 515.0\\
\hline\hline
\end{tabular}
\vspace{0.3cm}
\caption{\label{tab1} Binding energies $B_3^{(n)}$ of the triton excited
states and their ratios at the critical pion mass.
The second and 4th columns show the results from chiral EFT
while the third and 5th columns show the results from contact EFT.
The dashes indicate entries that have not been calculated, while the
blank entries are not defined.
}
\end{center}
\end{table}
from chiral EFT while the third and 5th columns show the 
binding energies and ratios from contact EFT, respectively.
The dashes indicate entries that have not been calculated, while the
blank entries are not defined.
{}From the ratios of the chiral EFT results it is evident that the
first two excited states are still influenced by higher-order
effects such as the finite range of the chiral potential. 
The exact ratio of 515.035... will be reached for the shallower states,
but their calculation in the chiral EFT is computationally very
expensive. The contact EFT, on the other hand, has an limit cycle 
in the ultraviolet by construction. If the theory is tuned to the critical
point, the limit cycle is exact for all energies. The 
ratio in the 5th column of Table \ref{tab1} is 
therefore 515.035... for all states. The contact theory
is most accurate for the shallower states where the chiral EFT also
approaches a limit cyle and becomes less accurate for the deeper
states. Note also that this theory predicts infinitely many deeper
states whose binding energies are beyond the range of validity of this
EFT. As an example, we have shown the state with $n=-1$ in Table \ref{tab1}.

The binding energies of the excited states directly at the critical point
can also be obtained analytically from the formula \cite{Efi71}
\beqa
B_3^{(n)} &=& \left(e^{2\pi/s_0}\right)^{1-n}\; B_3^{(1)}
= (515.035..)^{1-n}\;  B_3^{(1)}\,,
\eeqa
and the numerical results of the contact EFT are in good agreement with 
this analytical formula. 

If one is interested in small deviations from the exact limit cycle,
for example to calculate observables in the critical region,
precise numerical techniques are required. Results with 5 digits of 
precision as in this work can be obtained in a straighforward way.
Higher numerical accuracy requires more advanced techniques as described in 
Ref.~\cite{Mohr:2005pv}, where three-body binding energies 
up to 12 digits of precision have been obtained for the bosonic problem.
Moreover, the leading dependence of the three-body
energies on the physical ultraviolet cutoff, provided by the long-range
one-pion exchange in our case, was also given in Ref.~\cite{Mohr:2005pv}:
A renormalization group analysis suggests that the leading corrections
to the three-body binding energies $B_3^{(n)}$ are of order
$\sqrt{B_2} \ln(\Lambda)/\Lambda$ and  $B_3^{(n)}\ln(\Lambda)/\Lambda^2$
relative to $B_3^{(n)}$ itself. The first correction 
proportional to $\sqrt{B_2}$ vanishes at the critical
point. The second correction is qualitatively consistent with the results 
for the chiral EFT in Table \ref{tab1} and predicts a decreased
value for $B_3^{(2)}/B_3^{(3)}$. However, the decrease is too strong and 
would lead to a ratio $B_3^{(2)}/B_3^{(3)}<515$, so other corrections
must be important.

\begin{figure}[tb]
\centerline{\includegraphics*[width=9cm,angle=0,clip=true]{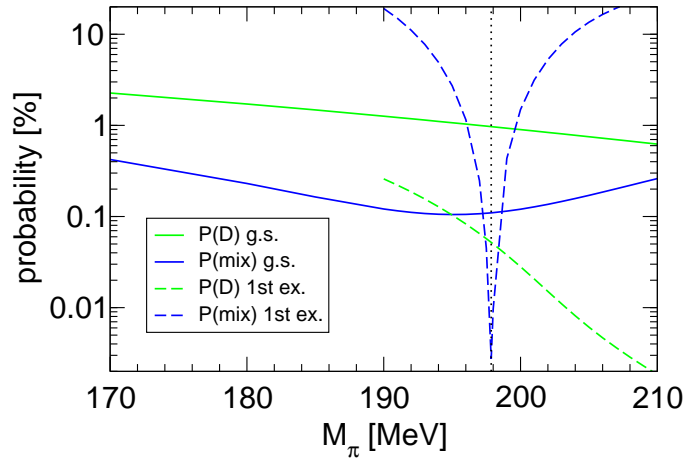}}
\caption{Probability of the D-wave and mixed symmetry states
for the triton ground state and the first excited state
as function of pion mass $M_\pi$.
The vertical dotted line indicates the critical pion mass $\mpic$.
}
\label{fig:Pdmix}
\end{figure}

It is interesting to compare the probability for the D-wave and 
mixed symmetry states in the chiral EFT with the expectations from 
the contact EFT. In Fig.~\ref{fig:Pdmix}, we show the
probability for the D-wave and mixed symmetry states
in the triton ground state and the first excited state
as function of pion mass $M_\pi$ for the chiral EFT.
The probabilities for the second excited state have also been 
calculated but are not shown. They behave similar to the first excited state.
The mixed symmetry probabilities
in the triton excited states become very small near the 
critical pion mass. This is a consequence of the $SU(4)$ symmetry at the 
critical point, which leads to a decoupling of the mixed symmetry state
\cite{Bedaque:1999ve}.

In the chiral EFT, the $SU(4)$ symmetry at the critical point
is exactly realized for the scattering lengths, but not 
for the effective ranges. This suggests that 
the mixed symmetry component is induced in the contact EFT
by higher order terms proportional to $r_e \sqrt{ME}$.
The observed suppression of the mixed symmetry 
component at the critical point is indeed consistent with this result 
of the power counting. The probabilities decrease roughly by factors of 30,
which is in the order of magnitude expected from the binding energy ratios 
of the states. The strong enhancement of the mixed symmetry 
state probabilities once the pion mass is tuned away from the critical point 
is generated by SU(4) breaking of the scattering lengths, which is a 
leading-order effect in the contact EFT. 
We observe that the mixed symmetry state is very large,
for the excited states. Treating SU(4) breaking interactions as a 
perturbation, this behavior might be explained by the proximity of other 
(virtual) excited states, which leads to an enhancement of the perturbative 
contributions to the wave function.   

For the D-state probabilities, similar observations hold. However, since 
the D-state component is generated by subleading interactions for all
pion masses, we do not observe the strong enhancement, once the 
pion mass is tuned away from the critical mass.  
All D-state probabilities decrease slowly with increasing pion
mass. This indicates that the tensor force becomes less important 
at larger pion masses. This can be expected, since in the limit 
of very large pion masses, the pion becomes a heavy degree of 
freedom, so that the approximation by contact interactions 
becomes more and more accurate.

\begin{figure}[tb]
\centerline{\includegraphics*[width=9cm,angle=0,clip=true]{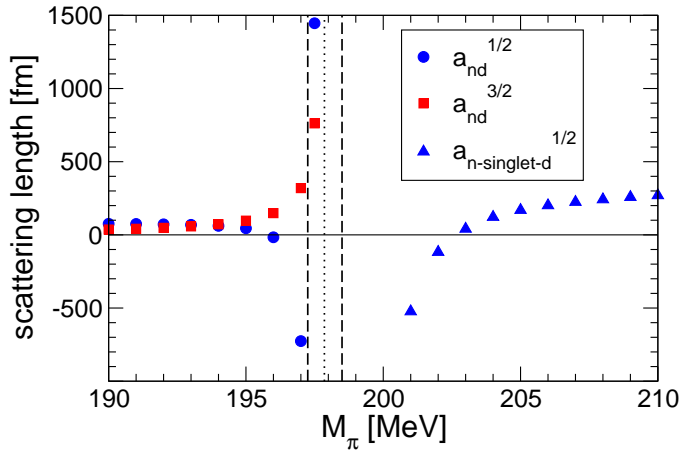}}
\caption{The $S=1/2$ and $S=3/2$ neutron-deuteron 
scattering lengths ($M_\pi \leq \mpic$)
and the $S=1/2$ neutron-singlet-deuteron scattering length 
($M_\pi \geq \mpic$) as function of pion mass $M_\pi$. 
The vertical dotted line indicates the critical pion mass $\mpic$
while the vertical dashed lines show where the second excited state
appears in the spectrum.
}
\label{fig:A1232}
\end{figure}

The calculation of three-body scattering observables 
in the contact EFT is also straightforward \cite{Bedaque:2002mn}.
In Fig.~\ref{fig:A1232}, we show results for the  
$S=1/2$ and $S=3/2$ neutron-deuteron scattering lengths 
for $M_\pi \leq \mpic$. Above the critical pion mass the deuteron ceases 
to exist, but the spin-singlet-deuteron becomes bound.
For $M_\pi \geq \mpic$, we also show the 
$S=1/2$ neutron-singlet-deuteron scattering length.
The vertical dashed lines show where the second excited state
appears in the spectrum. 
The critical point is indicated by the vertical dotted line.

All scattering lengths diverge at the critical point. The $S=3/2$
scattering length 
diverges because it is simply a constant times the spin-triplet 
scattering length: $1.179\,a_{\trip}$ \cite{STM57}.
The $S=1/2$ scattering lengths are very sensitive to the appearance
of new excited states at threshold as the  inverse two-body scattering 
lengths approach zero. They jump from $-\infty$
to $+\infty$ whenever a new state appears. 

The possible existence of a limit cycle in the pion mass region around 
200 MeV must also be taken into account in future nuclear lattice 
calculations. If the pion mass is in the critical region, 
three-body scattering observables will show a strong pion mass dependence. 
Since, however,
the physics of the limit cycle can be captured using EFT methods
as demonstrated in this work, it should not render
such calculations impossible. One can simply rely on the EFT to 
extrapolate through the critical region to physical pion masses.
A similar strategy was followed in a recent fully dynamical
lattice calculation of the nucleon-nucleon scattering lengths \cite{Beane06}.

\section{Summary \& Conclusion}

In this paper, we have performed the first study of the conjectured
infrared limit cycle in QCD in the framework of chiral EFT. 
We have exemplified three parameter sets in the chiral EFT at NLO 
that lead to an infrared limit cycle for pion masses around 200 MeV
and investigated the structure of three-body 
observables for one of these sets in detail. 

Using both chiral and contact EFT's, we have calculated 
the energies and structure of the triton ground state and the first 
two excited states around the critical pion mass.
Furthermore, we have calculated the next four excited states and 
the neutron-deuteron and neutron-singlet-deuteron scattering lengths
in the critical region. All three-body scattering lengths diverge
at the critical point. The $S=1/2$ scattering lengths are very sensitive 
to the appearance of new excited states in the critical region.

Moreover, we have elucidated the consequences for future
three-body lattice QCD calculations. On the one hand, scattering observables
are very sensitive to new three-body states appearing at threshold in the
critical region. On the other hand, the energies of the three-body states 
themselves and the triton ground state energy in particular change only
very slowly as one passes through the critical region. This suggests that 
chiral EFT could be a reliable tool for extrapolations of lattice results 
from unphysically large pion masses to the physical value.

The comparison of the chiral and contact EFT results also 
showed that an accurate matching of both theories is possible
around the critical point. The small deviations of the contact 
EFT predictions from the full chiral result were in line 
with the expectations from the power counting of the contact EFT.  
It is reassuring to confirm these expectations by an explicit calculation. 

In summary, the main findings of Ref.~\cite{Braaten:2003eu} have
been confirmed and extended. Future lattice studies promise 
interesting insights into the rich structure of QCD. In particular,
it would be very interesting to see signatures of the limit cycle 
in lattice simulations of the three-nucleon system
with pion masses around 200 MeV \cite{Wilson:2004de}.

\begin{acknowledgments}
This work was supported by the U.S. Department of Energy  
Contract No. DE-AC05-84ER40150 under which the Southeastern
Universities Research Association (SURA) operates the The Thomas Jefferson
National Accelerator Facility,
the EU Integrated Infrastructure Initiative Hadron Physics
under contract number RII3-CT-2004-506078,
and the Deutsche Forschungsgemeinschaft through funds provided
to the SFB/TR 16 \lq\lq Subnuclear structure of matter''. 
The numerical calculations have partly been performed on the JUMP 
cluster of the NIC, J\"ulich.
\end{acknowledgments}

 
\end{document}